\documentclass[11pt]{article}

\usepackage{amssymb}
\usepackage{citesort}

\begin{document}

\begin{flushright}
Preprint TPJU--12/03 \\
hep-th/0312179
\end{flushright}
\bigskip

\begin{center}

{\LARGE\bf\sf The square root of the Dirac operator on the superspace \\

\vskip 3mm

and the Maxwell equations}
\end{center}
\bigskip

\begin{center}
    {\large\bf\sf
    Adam Bzdak\footnote{e-mail: bzdak@th.if.uj.edu.pl}
    and
    Leszek Hadasz\footnote{e-mail: hadasz@th.if.uj.edu.pl}} \\
\vskip 3mm

    M. Smoluchowski Institute of Physics, \\
    Jagiellonian University,
    Reymonta 4, 30-059 Krak\'ow, Poland
\end{center}

\vskip .5cm

\begin{abstract}
We re-consider the procedure of ``taking a square root of the
Dirac equation'' on the superspace and show that it leads to the
well known superfield $W_\alpha$ and to the proper equations of
motion for the components, {\em i.e.} the Maxwell equations and
the massless Dirac equation.
\end{abstract}


\vspace*{2cm}


\section{Introduction}
The Dirac equation is often ``derived'' by requiring that it squares
to the Klein-Gordon equation \cite{Dirac}. In this sense one can view the Dirac
operator as a square root of the Klein-Gordon operator. It is then
natural to ask if this procedure can be taken one step further, {\em i.e.}
if one can give a sensible and useful definition of the square root of the
Dirac operator.

Such a construction within a ${\cal N} = 1$ superspace was given in
\cite{Szwed:1986ne} and extended to ${\cal N} = 2$ case in
\cite{Ketov:cp}. The main conclusion that followed
(expressed implicitly in \cite{Szwed:1986ne} and quite
explicitly in \cite{Ketov:cp}) was a bit disappointing: the equations
that result from taking a square root of the Dirac operator are just
constraints imposed on the superfield which eliminates some of the
components, but do not yield the dynamics to the remaining ones.

The main goal of this letter is to re-analyze the situation
and to show that (providing one requires that the obtained equations
square to the Dirac equation and not to its ``generalization'')
such a conclusion is not correct: properly understood Szwed equations
lead to dynamical equations for the component fields. We discuss in
details the simplest case, with the solution to the Szwed equations
being the Maxwell superfield with the vector component (the ``photon'')
satisfying the usual Maxwell equations and the spinor component
(the ``photino'') satisfying the massless Dirac equation.

\section{The square root of the Dirac operator}
Using the two-component notation and the chiral representation for
the Dirac matrices (for the conventions see \cite{WessAndBagger})
one can write the Dirac equation as
\begin{equation}
\label{Dirac}
-\left(
\begin{array}{cc}
i\bar\sigma^{\mu\,\dot\alpha\alpha}\partial_\mu & m \\
m & i\sigma^\mu{}_{\alpha\dot\alpha}\partial_\mu
\end{array}
\right)
\left(
\begin{array}{c}
\varphi_\alpha \\ \bar\chi^{\dot\alpha}
\end{array}
\right)
\; \equiv \;
{\cal D}\left(
\begin{array}{c}
\varphi_\alpha \\ \bar\chi^{\dot\alpha}
\end{array}
\right)
\; = \; 0.
\end{equation}
We are looking for an operator $A$ which satisfies
\begin{equation}
\label{prop1}
A^\dag\!A = {\cal D}.
\end{equation}
If one requires $A$ to be local and to contain space-time derivatives,
then the only possibility of satisfying (\ref{prop1}) is to allow $A$ to
depend on the anticommuting variables. Indeed there are no
second order space-time derivatives in the Dirac operator,
and one can check that the coefficient at $\partial_\mu$ in $A$ has to square to zero.
We are therefore lead to consider the operator $A$ as acting on the superspace
with the coordinates $(x^\mu, \theta^\alpha,\bar\theta^{\dot\alpha}).$

A solution to (\ref{prop1}) is given by
\begin{equation}
\label{A:solution}
A \; = \;
{1\over\sqrt 2}
\left(
\begin{array}{rr}
D^\alpha & -\bar D_{\dot\alpha} \\
\bar D^{\dot\alpha} & D_\alpha
\end{array}
\right)
\end{equation}
where the differential operators
\begin{eqnarray}
D_\alpha & = & \;\;\partial/{\partial\theta^\alpha} +
i\sigma^\mu{}_{\alpha\dot\alpha}\bar\theta^{\dot\alpha}\partial_\mu,
\nonumber \\
\bar D_{\dot\alpha} & = & \!-{\partial}/{\partial\bar\theta^{\dot\alpha}}
-i\theta^{\alpha}\sigma^\mu{}_{\alpha\dot\alpha}\partial_\mu,
\end{eqnarray}
satisfy the algebra
\begin{eqnarray}
\label{algebra}
\left\{D_\alpha,D_\beta\right\}
& = &
\left\{\bar D_{\dot\alpha},\bar D_{\dot\beta}\right\}
= 0,
\nonumber \\
\left\{D_\alpha,\bar D_{\dot\beta}\right\}
& = & -2i\sigma^\mu{}_{\alpha\dot\beta}\partial_\mu.
\end{eqnarray}

From (\ref{A:solution}) we get
\begin{equation}
\label{A:explicit}
A^\dag\!A =
-\left(
\begin{array}{cc}
i\bar\sigma^{\mu\dot\alpha\alpha}\partial_\mu & M \\
M & i\sigma^\mu{}_{\alpha\dot\alpha}\partial_\mu
\end{array}
\right)
\end{equation}
with
\begin{equation}
\label{M:definition}
M \; = \; -\frac14\left(D\!D +\bar D\!\bar D\right).
\end{equation}

We conclude that the operators $D$ and $A^\dag A$ are equal on the space of fields $F$
which satisfy
\begin{equation}
\label{additional}
\left(D\!D +\bar D\!\bar D\right)F + 4m\,F = 0.
\end{equation}
On this space we can therefore consider the equation
\begin{equation}
\label{A:equation}
A\,F = 0
\end{equation}
as a square root of the Dirac equation.

\newpage
The spinorial derivatives $D_\alpha,\bar D_{\dot\alpha}$
anticommute with the supersymmetry generators
\begin{equation}
\label{anticoms}
\left\{D_\alpha,Q_\beta\right\}
=
\left\{D_\alpha,\bar Q_{\dot\beta}\right\}
 =
\left\{\bar D_{\dot\alpha},Q_\beta\right\}
=\left\{\bar D_{\dot\alpha},\bar Q_{\dot\beta}\right\}
= 0.
\end{equation}
Consequently, the SUSY transformations generated by $Q_\alpha,\bar Q_{\dot\alpha},$
\[
\delta F = (\epsilon Q + \bar\epsilon \bar Q)F,
\]
where $\epsilon$ is an infinitesimal, constant, anticommuting spinor,
transform one solution of the system of equations (\ref{additional}), (\ref{A:equation})
into another.

\section{A solution}
We shall be interested in the solutions of
(\ref{A:equation}) having the form\footnote{
We shall exclude another form of solution discussed in \cite{Szwed:1986ne} and \cite{Ketov:cp}
\[
B\;=\;\left(
\begin{array}
[c]{c}%
\Phi\;\\
V^{\alpha\dot{\alpha}}%
\end{array}
\right)\label{B}%
\]
from our considerations, as the equation $A^\dag\!A\,B = 0$
is not equivalent to the Dirac equation.}
\begin{equation}
\label{F}
F \; = \;
\left(
\begin{array}{r}
W_\alpha \\
\bar{\cal H}^{\dot\alpha}
\end{array}
\right).
\end{equation}
Using (\ref{A:solution}) we get
\begin{eqnarray}
\label{explicit}
D^\alpha W_\alpha - \bar D_{\dot\alpha}\bar{\cal H}^{\dot\alpha} & = & 0,
\nonumber \\
\bar D^{\dot\alpha} W_\alpha +D_\alpha\bar{\cal H}^{\dot\alpha} & = & 0.
\end{eqnarray}

Let us now work out some of the consequences of (\ref{explicit}) and (\ref{additional}).

It is not difficult to show that $W_\alpha$ and $\bar{\cal H}^{\dot\alpha}$
are a fully fledged superfields, {\em i.e.} their components at the non-zero powers
of $\theta_\alpha , \bar\theta^{\dot\alpha}$ are non-vanishing. Indeed if we assume
that
\[
W_\alpha = \lambda_\alpha(x),
\hskip 1cm
\bar{\cal H}^{\dot\alpha} = \bar h^{\dot\alpha}(x),
\]
then it follows from (\ref{explicit}) that
\[
\partial_\mu\lambda_\alpha = \partial_\mu \bar h^{\dot\alpha} = 0,
\]
so the only allowed solutions are constants.

Acting with  the $D^\alpha$ derivative on the equations (\ref{explicit})
and using the equality
\begin{equation}
\label{Dirac2}
i\bar\sigma^{\mu\,\dot\alpha\alpha}\partial_\mu W_\alpha +
M\bar{\cal H}^{\dot\alpha} = 0,
\end{equation}
(which follows from (\ref{A:explicit}) and the fact that $AF = 0\; \Rightarrow A^\dag\!A\, F= 0$),
we get
\begin{equation}
\label{secondorder:a}
D^2\bar{\cal H}^{\dot\alpha} = 0.
\end{equation}
Similarly one shows that
\begin{equation}
\label{secondorder:b}
\bar D^2W_\alpha = 0.
\end{equation}
After some more algebra one checks that the relations
\[
\partial_\mu\left(\bar D^{\dot\alpha} W_\alpha\right)
=
D^\beta\left(\bar D^{\dot\alpha} W_\alpha\right)
=
\bar D^{\dot\beta}\left(\bar D^{\dot\alpha} W_\alpha\right)
=0
\]
and
\[
\partial_\mu\left(D_\alpha\bar{\cal H}^{\dot\alpha}\right)
=
D^\beta\left(D_\alpha\bar{\cal H}^{\dot\alpha}\right)
=
\bar D^{\dot\beta}\left(D_\alpha\bar{\cal H}^{\dot\alpha}\right)
=0
\]
hold, what proves the equality
\begin{equation}
\label{chiral1}
\bar D^{\dot\alpha} W_\alpha
=
-D_\alpha\bar{\cal H}^{\dot\alpha}
=
{\rm const.}
\end{equation}
Let us denote this constant by $\eta^{\dot\alpha}_\alpha$ and notice
that (\ref{additional}) and (\ref{explicit}) are invariant under the shift
\begin{eqnarray}
\label{shift}
W_\alpha & \to & W_\alpha +\bar\theta_{\dot\alpha}\eta^{\dot\alpha}_\alpha,
\nonumber \\
\bar{\cal H}^{\dot\alpha}
& \to &
\bar{\cal H}^{\dot\alpha} -\theta^\alpha\eta^{\dot\alpha}_\alpha.
\end{eqnarray}
The shifted fields thus satisfy
\begin{equation}
\label{chiral2}
\bar D^{\dot\alpha} W_\alpha
= 0,
\hskip 1cm
D_\alpha\bar{\cal H}^{\dot\alpha}
=0.
\end{equation}
It is worth to emphasize that the result above does not depend on the equation (\ref{additional}).

It is intriguing to notice that even if we started from the massive Dirac equation,
consistency requires $m$ to be equal to zero.
Indeed it follows from (\ref{additional}) that $W_\alpha$ satisfies
\begin{equation}
\label{eigeneq}
(DD + \bar D\bar D)W_\alpha = -4mW_\alpha
\end{equation}
Acting on this equation with $D^\alpha$ we get (using (\ref{secondorder:b}))
\[
-4mD^\alpha W_\alpha = DDD^\alpha W_\alpha = 0.
\]
If we assume that $m \neq 0$ then acting with $D^\alpha$ on the equation
(\ref{chiral1}), we get
\[
\bar\sigma^{\mu\dot\alpha\alpha}\partial_\mu W_\alpha = 0
\]
so that from (\ref{Dirac2}):
\[
m \bar{\cal H}^{\dot\alpha} = 0
\hskip 5mm
\Rightarrow
\hskip 5mm
\bar{\cal H}^{\dot\alpha} = 0.
\]
Similarly one can show that the assumption $m \neq 0$ leads to $W_\alpha = 0.$ Therefore, in order
to obtain non-trivial solutions to the Szwed equation
(\ref{A:equation}) one has to assume that
\begin{equation}
\label{mass0}
m = 0.
\end{equation}

Let us now consider the simplest case $W_\alpha = {\cal H}_\alpha.$
Using  (\ref{chiral2}) one can see that equations (\ref{explicit})
now read
\begin{equation}
\label{constr}
\bar D^{\dot\alpha} W_\alpha \; = \; 0,
\hskip 1cm
D_\alpha\bar{W}^{\dot\alpha}\; = \; 0,
\hskip 1cm
D^\alpha W_\alpha = \bar D_{\dot\alpha}\bar{W}^{\dot\alpha}.
\end{equation}
It follows \cite{WessAndBagger,Weinberg} that $W_\alpha$ has the form
of the Maxwell superfield,
\begin{eqnarray}
\label{structure}
W_\alpha & = & -i\lambda_\alpha(y) + \left[\delta_\alpha{}^\beta d(y)
-\frac{i}{2}\left(\sigma^\mu\bar\sigma^\nu\right)_\alpha{}^\beta F_{\mu\nu}(y)\right]\theta_\beta
+\theta\theta\sigma^\mu{}_{\alpha\dot\alpha}\partial_\mu\bar\lambda^{\dot\alpha}(y).
\end{eqnarray}
Here $F^{\mu\nu}$ is an antisymmetric tensor,
$y^\mu = x^\mu + i\theta^\alpha\sigma^\mu{}_{\alpha\dot\alpha}\bar\theta^{\dot\alpha},$
\begin{equation}
d^\ast(x) = d(x),
\hskip 1cm
F^{\ast\mu\nu}(x) = F^{\mu\nu}(x),
\end{equation}
(star denotes complex conjugation) and
\begin{equation}
\label{Bianchi}
\varepsilon^{\mu\nu\rho\lambda}\partial_\nu F_{\rho\lambda} = 0.
\end{equation}

Now we may employ equation (\ref{eigeneq}) which, in view of (\ref{secondorder:b})
and (\ref{mass0}) takes the form
\begin{equation}
\label{masseqsimple}
D\!D\,W_\alpha = 0.
\end{equation}
From  (\ref{masseqsimple}) and its complex conjugation one can check
(taking steps similar to those that lead to the equality (\ref{chiral1}))
that the relations
\[
\partial_\mu\left(D^{\alpha} W_\alpha\right)
=
D^\beta\left(D^{\alpha} W_\alpha\right)
=
\bar D^{\dot\beta}\left(D^{\alpha} W_\alpha\right)
=0
\]
hold, what proves the equality
\begin{equation}
\label{almostfinal}
D^\alpha W_\alpha = \bar D_{\dot\alpha}\bar{W}^{\dot\alpha} = {\rm const.}
\end{equation}
For $W_\alpha$ given in (\ref{structure}) this boils down to the equations
for the component fields of the form
\begin{equation}
\label{final1}
d = {\rm const.},
\hskip 5mm
\sigma^\mu{}_{\alpha\dot\alpha}\partial_\mu\bar\lambda^{\dot\alpha} = 0,
\end{equation}
and
\begin{equation}
\label{final2}
\partial_\nu F^{\nu\mu} = 0.
\end{equation}
The equations (\ref{final2}) and (\ref{Bianchi}) form a full set
of Maxwell equations for the antisymmetric, real tensor $F^{\mu\nu}.$

It may be interesting to extend the results from this letter in
several directions.

Starting from the Dirac equation on the
curved space--time one may hope to end up with the equations
covariant with respect to the local supersymmetry transformations,
``re-deriving'' in this way some simple supergravity model.

It is known that by ``taking a square root'' of the Klein--Gordon operator one can
obtain not only the Dirac equation, but also the first order equations
describing the massive scalar or the massive vector fields (the Dirac matrices
are in this case replaced by the matrices satisfying the Duffin--Kemmer algebra).
It is tempting to try to repeat this
procedure here and to check if one arrives at the equations describing other supermultiplets.

Last but not least, one may try to ``take a square root'' of the
coupled Maxwell--Dirac equations. With a bit of luck this may lead
to an interacting supersymmetric model.

\end{document}